\documentclass[5p, twocolumn]{elsarticle}
\usepackage{color} 
\usepackage{tabularx} 
\usepackage{amsmath} 
\usepackage{amssymb} 
\usepackage{graphicx}
\usepackage{wasysym}
\usepackage{times}
\usepackage{hyperref}
\usepackage{booktabs,multirow}
\usepackage{xurl}

\begin{document}
\begin{frontmatter}

\title{Computational Overhead of Locality Reduction in Binary Optimization Problems}

\author[1]{Elisabetta Valiante\corref{cor1}%
	}
\ead{Elisabetta.Valiante@1qbit.com}

\author[1]{Maritza Hernandez}
\ead{Maritza.Hernandez@1qbit.com}

\author[2]{Amin Barzegar}
\ead{Amin.Barzegar@microsoft.com}

\author[3,4,5]{Helmut G.~Katzgraber\fnref{fn1}}
\ead{katzgrab@amazon.com}

\cortext[cor1]{Corresponding author}
\fntext[fn1]{ The work of H.~G.~K.~was performed before joining Amazon Web Services.}

\affiliation[1]{organization={1QB Information Technologies (1QBit)}, 
		addressline={1285 W Pender St Unit 200}, 
		city={Vancouver}, 
		postcode={British Columbia V6E~4B1}, 
		country={Canada}}
\affiliation[2]{organization={Microsoft Quantum, Microsoft}, 
		city={Redmond}, 
		postcode={Washington 98052}, 
		country={USA}}
\affiliation[3]{organization={Amazon Quantum Solutions Lab}, 
		city={Seattle}, 
		postcode={Washington 98170}, 
		country={USA}}		
\affiliation[4]{organization={AWS Intelligent and Advanced Compute Technologies, Professional Services}, 
		city={Seattle}, 
		postcode={Washington 98170}, 
		country={USA}}
\affiliation[5]{organization={AWS Center for Quantum Computing}, 
		city={Pasadena}, 
		postcode={California 91125}, 
		country={USA}}
\date{\today}

\begin{abstract}
Recently, there has been considerable interest in solving optimization problems by mapping these onto a binary representation, sparked mostly by the use of quantum annealing machines. Such binary representation is reminiscent of a discrete physical two-state system, such as the Ising model. As such, physics-inspired techniques---commonly used in fundamental physics studies---are ideally suited to solve optimization problems in a binary format. While binary representations can be often found for paradigmatic optimization problems, these typically result in $k$-local higher-order unconstrained binary optimization cost functions. In this work, we discuss the effects of locality reduction needed for the majority of the currently available quantum and quantum-inspired solvers that can only accommodate $2$-local (quadratic) cost functions. General locality reduction approaches require the introduction of ancillary variables which cause an overhead over the native problem. Using a parallel tempering Monte Carlo solver on Microsoft Azure Quantum, as well as $k$-local binary problems with planted solutions, we show that post reduction to a corresponding $2$-local representation the problems become considerably harder to solve. We further quantify the increase in computational hardness introduced by the reduction algorithm by measuring the variation of number of variables, statistics of the coefficient values, and the population annealing entropic family size. Our results demonstrate the importance of avoiding locality reduction when solving optimization problems.
\end{abstract}


\end{frontmatter}

\section{Introduction}
\label{sec:introduction}

In recent years, there have been many technological and algorithmic advances when solving optimization problems, in an industrial setting in particular. Sparked by the work of \mbox{D-Wave Systems Inc.} \cite{johnson:11, dickson:13, bunyk:14}, a whole new field of optimization based on physical processes has emerged. Some examples are: digital processors based on simulated annealing \cite{yamaoka:17, yamamotok:17, tsukamoto:17, aramon:19} or other recently proposed algorithms \cite{okuyama:19,patel:20,yamamotok:21,leleu:21}; coherent Ising machines implemented with pulse lasers \cite{wang:13, marandi:14, inagaki:16, mcmahon:16, yamamotoy:17, hamerly:19}, and other kinds of optical Ising machine \cite{pierangeli:19, pierangeli:20a, pierangeli:20b, pierangeli:21}; simulated bifurcation \cite{goto:19b, goto:21} and other optimization using nonlinear oscillation networks \cite{goto:16, nigg:17, puri:17, goto:18, goto:19a}. Specifically, the development of hardware quantum annealers has stimulated new ways of tackling NP-hard problems previously inaccessible. 

Despite these advances, the use of quantum annealers for large-scale industry applications remains limited if not paired with classical algorithms on CMOS hardware. Being able to tackle an application requires first having a Boolean representation of the problem. To this {\em mapping} step, in most cases a variable overhead is associated, which typically makes a problem harder to solve. However, due to hardware limitations, only $2$-local (quadratic unconstrained binary optimization, or QUBO) cost functions can be tackled with quantum annealing hardware. This means that a higher-order binary polynomial unconstrained optimization problem requires a {\em locality reduction} which can result in a sizable variable overhead \cite{perdomo_ortiz:19}. In this work, we focus on the {\em locality reduction}, and do not discuss additional overheads due to the {\em embedding} of a binary problem onto the hardwired sparse quasi-two-dimensional topology of annealing hardware or the effects of analog noise. See Refs.~\cite{perdomo_ortiz:19, konz:21} and Refs.~\cite{zhu:16, albash:19} for analyses of overheads introduced by topology and analog noise, respectively. 

The hardware limitations play an important role when solving problems naturally formulated as a Hamiltonian with \mbox{$k$-local} interactions with $k>2$. There are various optimization problems in fundamental physics, computer science, and applications that are natively $k$-local. Examples in physics, as well as computer science, are computing the partition function of a four-dimensional pure lattice gauge theory~\cite{de_las_cuevas:09, de_las_cuevas:10}, measuring the fault-tolerance in topological colour codes~\cite{andrist:11}, and solving $k$-SAT problems with $ k>2$. Examples of practical applications are circuit fault diagnosis~\cite{feldman:10, perdomo_ortiz:19}, molecular similarity measurement~\cite{hernandez:16}, molecular conformational sampling~\cite{marchand:19}, and traffic light synchronization~\cite{qmedia}.

Quadratization techniques are algorithms used to reduce a higher-degree multilinear polynomial into a quadratic one \cite{boros:14}. The reduction process can introduce two different types of overheads. First, the quadratization itself can result in a large overhead before any solver is applied to the problem of interest. Nevertheless, the process is known to scale in polynomial time \cite{boros:02}. Second, quadratization requires the introduction of additional variables and terms. As such, the complexity of the problem increases and, in turn, so does the time to solution. Finally, the quadratization process might also introduce features (e.g., broader coupler distributions) that can affect the intrinsic difficulty of the problem. Both types of overheads might increase the complexity of the problem to the point of affecting its scaling. An extensive comparison between several quadratization methods, highlighting the pros and cons of each method, has been compiled by Dattani in Ref.~\cite{dattani:19}.

In this paper, we use Microsoft Azure Quantum's $k$-local solvers based on simulated annealing and parallel tempering Monte Carlo to measure the time overhead introduced by the quadratization process to reduce an optimization problem with \mbox{$k$-local} interaction to its $2$-local counterpart. We study unconstrained problems with a binary representation and planted solutions and disregard the time it takes for the quadratization algorithm to run. Our results demonstrate that the locality reduction introduces a large overhead when solving the problems. Employing a commonly used proxy metric, we demonstrate that, on average, optimization problems become much harder to solve when the locality is reduced. Reference \cite{konz:21} studies the embedding overhead when using sparse hardware topologies. Both complementary studies highlight the importance of developing new optimization machines and techniques that can handle $k$-local cost functions natively on complete graphs. 

This paper has the following structure: in Sec.~\ref{sec:problems}, we describe the benchmark problems used for the experiment; in Sec.~\ref{sec:benchmark}, we present the setup of the experiment and the metrics used to compare performance; in Sec.~\ref{sec:results} and Sec.~\ref{sec:discussion}, we discuss and analyze the results of the experiment; in Sec.~\ref{sec:concl}, we present our conclusions.

\section{Benchmark Problems}
\label{sec:problems}

In order to study the computational overhead caused by reducing a $k$-local problem to a quadratic ($2$-local) formulation, we first generate Ising problems for $k=3$ and $k=4$. The $k$-local instances have been generated using the \texttt{Chook} package, which is publicly available on GitHub; see~\cite{perera:20chook}. Using this package, we are able to construct planted-solution instances, thus ensuring that the ground state and corresponding energy are known a priori. The construction of $k$-local problems is performed by combining tile planting problems of lower-order, that is, $k \leq 2$. 

The tile planting method ~\cite{hamze:18, perera:20} decomposes the problem graph into edge-disjoint vertex-sharing subgraphs. It produces scalable problems with highly tunable complexity. \texttt{Chook} supports the generation of tile-planted problems on square and cubic lattice topologies with periodic boundary conditions. The regular structure of these lattices allows for a problem-graph decomposition that naturally renders a subset of the unit cells as subgraphs. Each subgraph is associated with an Ising cost Hamiltonian, and their sum defines the complete Hamiltonian of the problem.

The square lattices used in this work are defined by four subproblem classes that correspond to unit cycles (plaquettes) with different levels of frustration. A subproblem is constructed by assigning to the couplers values equal to $-1$, $1$, or $2$, according to the class to which the subproblem belongs. The class is assigned with a certain probability, and each instance class is defined by three probability parameters. We set the probability parameters to the default values used in \texttt{Chook}.

\texttt{Chook} constructs higher-order $k$-local problems ($k > 2$) by combining $n$ Hamiltonians $\mathcal{H}^{(i)}$ with lower-order ($k \leq 2$) interactions and known ground states. If the Hamiltonians are completely independent of each other in that the underlying problem graphs do not share any vertices or edges, the composite Hamiltonian $\mathcal{H}_{\rm comp}$, obtained as the product of the $\mathcal{H}^{(i)}$, is minimized by any of the $n$ known ground states. The locality of $\mathcal{H}_{\rm comp}$ is given by $k_{\rm max}=\sum_{i=1}^n k_{\rm max}^{(i)}$, with $k_{\rm max}^{(i)}$ being the locality of the highest-order term in the $\mathcal{H}^{(i)}$.

In this study, $3$-local instances have been generated by combining a square tile planting problem with Ising spins coupled to a bimodal random field, while for $4$-local instances, the problems have been generated by combining two square tile planting problems. For each locality considered, we generate instances with problem sizes $N$ (number of variables) between $16$ and $400$.

The $k$-local instances are then reduced to their quadratic form using an iterative reduction-by-substitution algorithm~\cite{rosenberg:75, boros:02}. Here, we consider the terms in the problem with degree $k_t > 2$: we substitute the product of two binary variables with a new auxiliary variable and add a penalty term to enforce equality in the ground state. 

A simple example of reducing a term is the following. Let us assume we have a third-degree binary polynomial with a term $x_1x_2x_3$, where we substitute $y=x_1x_2$ and introduce a penalty term:
\begin{align}
    x_1x_2x_3 & \implies \nonumber\\
        &   yx_3+C_{x_1,x_2}(x_1x_2-2x_1y-2x_2y+3y).
\end{align}
The penalty term is always equal to $0$ when the value of the auxiliary variable $y$ is equal to the product of the binary variables $x_1$ and $x_2$. The constant $C_{x_1,x_2}$ ensures that the constraint associated with the substitution of the product $x_1x_2$ is always satisfied. In fact, the constraint has to be obeyed regardless of the value of the other terms of the polynomial.

When reducing our problem instances, this process is repeated until the final function becomes quadratic. Tuning the value of the constants $C_{x_a,x_b}$ is extremely important: a small value could return a $2$-local problem not having the same optimum as the original higher-order problem. Therefore, a large value is commonly used in various implementations of this algorithm. As suggested in Refs.~\cite{rosenberg:75, boros:02}, in a generic binary polynomial $P(x_1,x_2,\dots,x_n)=\sum_{i\in T} c_i t_i$, where $t_i$ are the terms of the polynomial, a single constant can be defined for any substitution by summing all the coefficients:
\begin{equation}
    C_{x_a,x_b}>\sum_{i\in T} |c_i|.
\end{equation}
The absolute values of the coefficients, in a large polynomial, can accumulate to a very large number: this can pose issues when attempting to solve problems on current analog quantum annealing hardware, because large coefficients amplify the effects of the analog noise. 

The reduction of $k$-local problems in this work is done via the \texttt{Hobo2Qubo} function available earlier through 1QBit's 1Qloud platform~\cite{1qloud}, which uses a tight bound for the penalty coefficient and sets it independently for each reduced term. The computational time required to reduce a single instance is negligible with respect to the time required by the solver. Moreover, the reduction from $k$-local to $2$-local is known to scale in polynomial time~\citep{boros:02}, that is, it should be negligible for large problem sizes, relative to the exponential scaling of the cost of solving the problem. 

The sizes and densities of the $2$-local instances obtained after reduction from $3$-local and $4$-local instances are shown in Tables~\ref{tb:3l} and \ref{tb:4l}, respectively. The number of variables increases considerably when reducing locality from $k$-local to $2$-local, as can be expected for a reduction-by-substitution algorithm.

The density of a $k$-local instance $\rho$ is calculated as the sum of the densities for each degree in the polynomial, normalized for the number of degrees $\geq 2$. The terms of the sum are calculated as the fraction of non-zero couplings over all the possible couplings for each degree. This is implemented as 
\begin{equation}\label{eq:density}
    \rho = \frac{1}{k-1}\sum_{k_{\rm t}=2}^{k}\frac{(N-k_{\rm t})!k_{\rm t}!}{N!}E_{k_{\rm t}}\,, 
\end{equation}
where $k$ is the locality of the polynomial. The sum is taken over all the degrees in the polynomial running from $k_{\rm t} = 2$ to $k_{\rm t} =k$, $E_{k_{\rm t}}$ is the number of individual terms with degree $k_{\rm t}$, and $N$ is the number of variables in the polynomial. For \mbox{$2$-local} instances, this expression is reduced to the common graph density expression. Tables~\ref{tb:3l} and \ref{tb:4l} report the mean densities calculated over 30 instances for each problem size. 

Notice that, for almost all problems, the densities decrease slightly after locality reduction. This is an interesting observation because, in general, a larger density is expected to be associated with larger complexity \cite{aramon:19}. 

\begin{table}[ht]
\caption{Reduction of $3$-local problems to $2$-local problems. Densities for each instance are calculated as per Eq.~\ref{eq:density}. The mean values (denoted by an overbar) and their standard deviations are calculated over the $30$ instances that have been generated. The number of variables of the reduced problems increases by a factor $\sim 3$.} 
\label{tb:3l}
\centering 
\begin{tabular}{rccccr}  
\hline
\hline 
\multicolumn{2}{c}{$3$-local} &&
\multicolumn{2}{c}{$2$-local reduction} \\ 
\hline 
\rule{0pt}{2.5ex}{\centering $N$} & {$\bar{\rho}$} && {$\bar{N}$} & {$\bar{\rho}$}  \\
$16$  & $0.568 \pm 0.020$  && $46.73 \pm 0.573$    & $0.329 \pm 0.009$ \\
$64$  & $0.398 \pm 0.014$ && $192.0$              & $0.295 \pm 0.003$ \\
$144$ & $0.364 \pm 0.008$ && $432.0$              & $0.294 \pm 0.002$ \\
$256$ & $0.352 \pm 0.007$ && $768.0$              & $0.294 \pm 0.002$ \\ 
$400$ & $0.345 \pm 0.005$ && $1200.0$             & $0.294 \pm 0.001$ \\
\hline
\hline 
\end{tabular}
\end{table}

\begin{table}[ht]
\caption{Reduction of $4$-local problems to $2$-local problems. Densities for each instance are calculated as per Eq.~\ref{eq:density}. The mean values (denoted by an overbar) and their standard deviations are calculated over the $30$ instances that have been generated. The number of variables of the reduced problems increases by a factor $\sim 6$.} 
\label{tb:4l}
\centering 
\begin{tabular}{rccccr}
\hline
\hline 
\multicolumn{2}{c}{$4$-local} &&
\multicolumn{2}{c}{$2$-local reduction} \\ 
\hline
\rule{0pt}{2.5ex}{\centering $N$} & {$\bar{\rho}$} && {$\bar{N}$} & {$\bar{\rho}$}  \\
$16$  & $0.615 \pm 0.023$ && $\;\;\;76.5 \pm 2.0$     & $0.301 \pm 0.013$ \\
$64$  & $0.295 \pm 0.017$ && $\;\;448.5 \pm 4.0$    & $0.167 \pm 0.004$ \\
$144$ & $0.210 \pm 0.008$ && $\;\;887.6 \pm 2.2$    & $0.176 \pm 0.002$ \\
$256$ & $0.179 \pm 0.007$ && $1501.3 \pm 3.6$   & $0.173 \pm 0.002$ \\ 
$400$ & $0.163 \pm 0.004$ && $2248.3 \pm 3.6$   & $0.174 \pm 0.001$ \\
\hline
\hline 
\end{tabular}
\end{table}

\section{Experiment Setup}
\label{sec:benchmark}

The simulations are performed using Microsoft Azure Quantum's solvers, which can handle $k$-local terms natively. There are two variants of the solvers, parameter-free solvers and standard solvers. The parameter-free version requires the user to enter only a timeout and automatically optimizes the parameters to find solutions to binary cost functions to high probabilities. The standard solvers instead require parameter optimization to obtain the optimal performance.

\subsection{Setup}
\label{sec:setup}

For the experiments, we use the parameter-free \texttt{ParallelTempering} (v1.0) solver \cite{pt}. The best values for temperatures, number of sweeps, and number of replicas are calculated internally and are customized for each submitted problem individually. At the time the experiments discussed in this manuscripts are performed (July 2020), the solver does not disclose the parameters chosen for the optimization. The only parameter to set is \texttt{timeout}, which is the time spent in the core solver loop (in seconds). It is worth specifying that \texttt{timeout} does not include the time spent by the solver to calculate the parameters that are used during the optimization process. The total time the solver needs to solve the problem is referred to as \texttt{runtime}. The advantage of using a parameter-free solver is that no tuning experiment is necessary. The disadvantage is that the \texttt{runtime} we measure includes both the time to calculate the parameters and the time to solve the problem. At the time of running the experiment, the parameters calculated by the solver are not returned to the user in the current implementation. As such, we cannot list them in this work. 
    
The benchmark experiment consists of solving 30 random instances for each system size and locality, as well as their respective $2$-local reduction (see Tables~\ref{tb:3l} and \ref{tb:4l} for details). For each of these instances, we perform 30 runs to gather statistics. We set \texttt{timeout} $=100$. In cases when $100$ is not enough time to find the ground-state energy, we increase \texttt{timeout} to $500$.

\subsection{Metrics}
\label{sec:metrics}

The primary objective of our benchmark experiment is to quantify how the computational effort in solving a problem scales as the size of the problem input increases. The common approach is to measure the time to solution (TTS). We calculate the TTS following the approach defined in Refs.~\cite{ronnow:14, aramon:19}:
\begin{equation}
    {\rm TTS}=\tau{\rm R}_\text{99},
\end{equation}
where ${\rm R}_{\text{99}}$ is the number of runs required to find the ground-state energy with a probability of $99\%$ and $\tau$ is the time it takes to run the algorithm once (i.e., the solver output \texttt{runtime}).

We derive ${\rm R}_{\text{99}}$ by estimating its distribution of the 50th percentile. This requires the algorithm to find the ground-state energy of each problem for at least $50\%$ of the successive runs performed (see Ref.~\cite{aramon:19} for more details). When it is not possible to measure the TTS, because the ground-state energy cannot be determined sufficiently often, we measure other performance metrics, such as the fraction of solved problems and the residual energies---both defined below. 

The fraction of solved problems is defined as the fraction of runs for which the ground-state energy is found by the solver divided by the total number of experiments. We have performed a total of 900 runs for each problem size and locality. The energy is calculated for each problem and each run in the following way:
\begin{equation}\label{eq:residuals}
    R = \frac{E_{\rm GS} - E_{\rm best}}{E_{\rm GS}},
\end{equation}
where $E_{\rm GS}$ is the known planted ground-state energy of the problem and $E_{\rm best}$ is the best energy found by the algorithm. The values reported here are obtained by resampling the distribution of residuals over all problems and runs.  

\section{Results}
\label{sec:results}

\begin{figure}[t!]
    \centering
    \includegraphics[width=\linewidth, keepaspectratio]{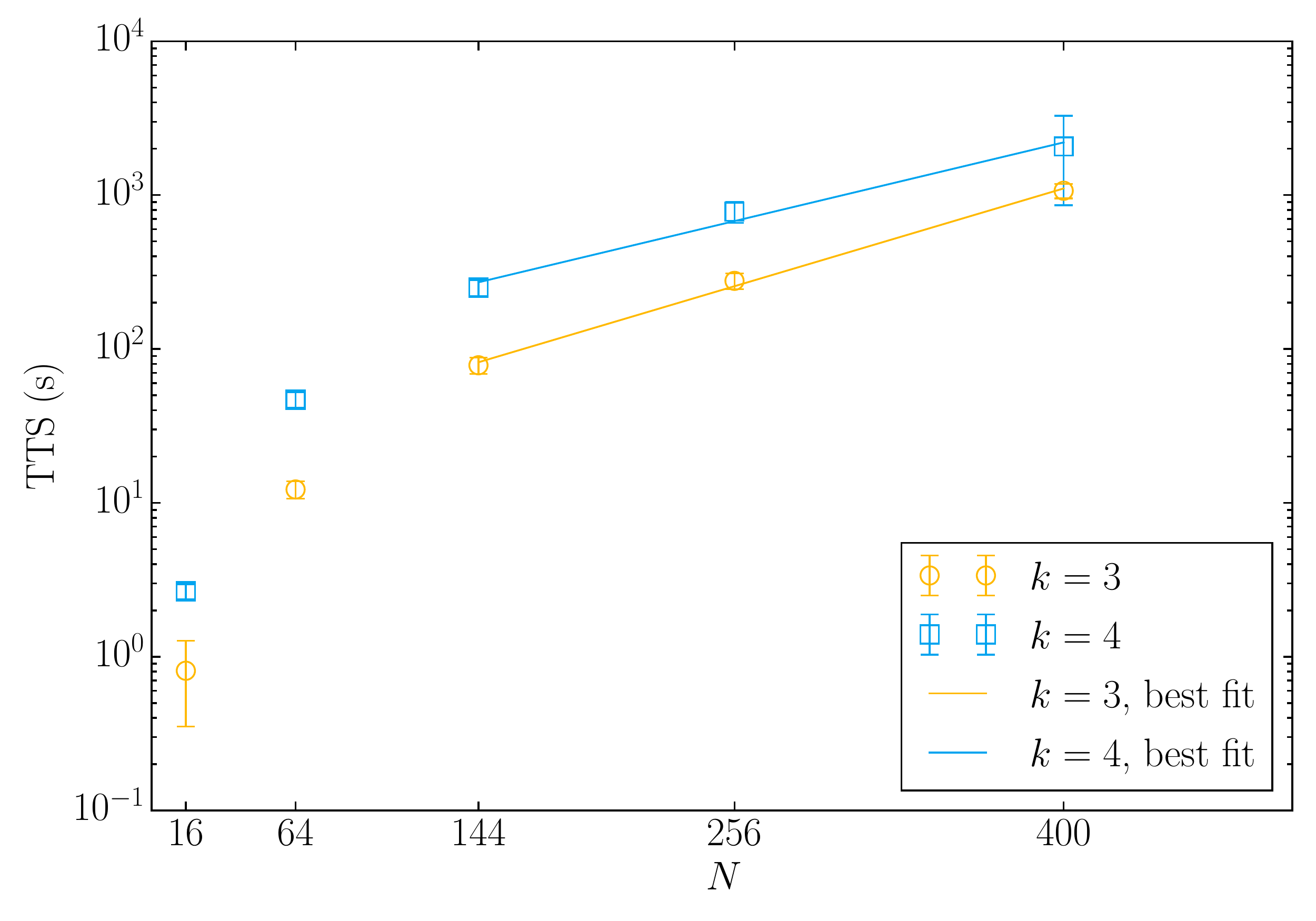}
    \caption{TTS mean value for $k$-local problems with $k=3$ and $k=4$ using the parallel tempering solver. The error bars correspond to a $2\sigma$ confidence interval. The continuous lines show the result of fitting an exponential function over the three largest problem sizes (see text for more details).}
    \label{fig:tts}
\end{figure}

Figure~\ref{fig:tts} shows the TTS for planted  $3$- and $4$-local problems with a number of variables $N$ ranging from $16$ to $400$ using the parallel tempering algorithm. Both problem types show a similar scaling. We have fit an exponential function of the form ${\rm TTS}=10^{\alpha+\beta N}$ over the three largest problem sizes. The results of the fit and the estimated scaling exponent $\beta$ are:
\begin{align*}
    \beta = 0.00441(14) \quad \quad (k=3)\\
    \beta = 0.00355(67) \quad \quad (k=4)
\end{align*}

\begin{figure}[t!]
    \centering
    \vspace{+1.2mm}
    \hspace*{+1mm}\includegraphics[width=0.96\linewidth, keepaspectratio]{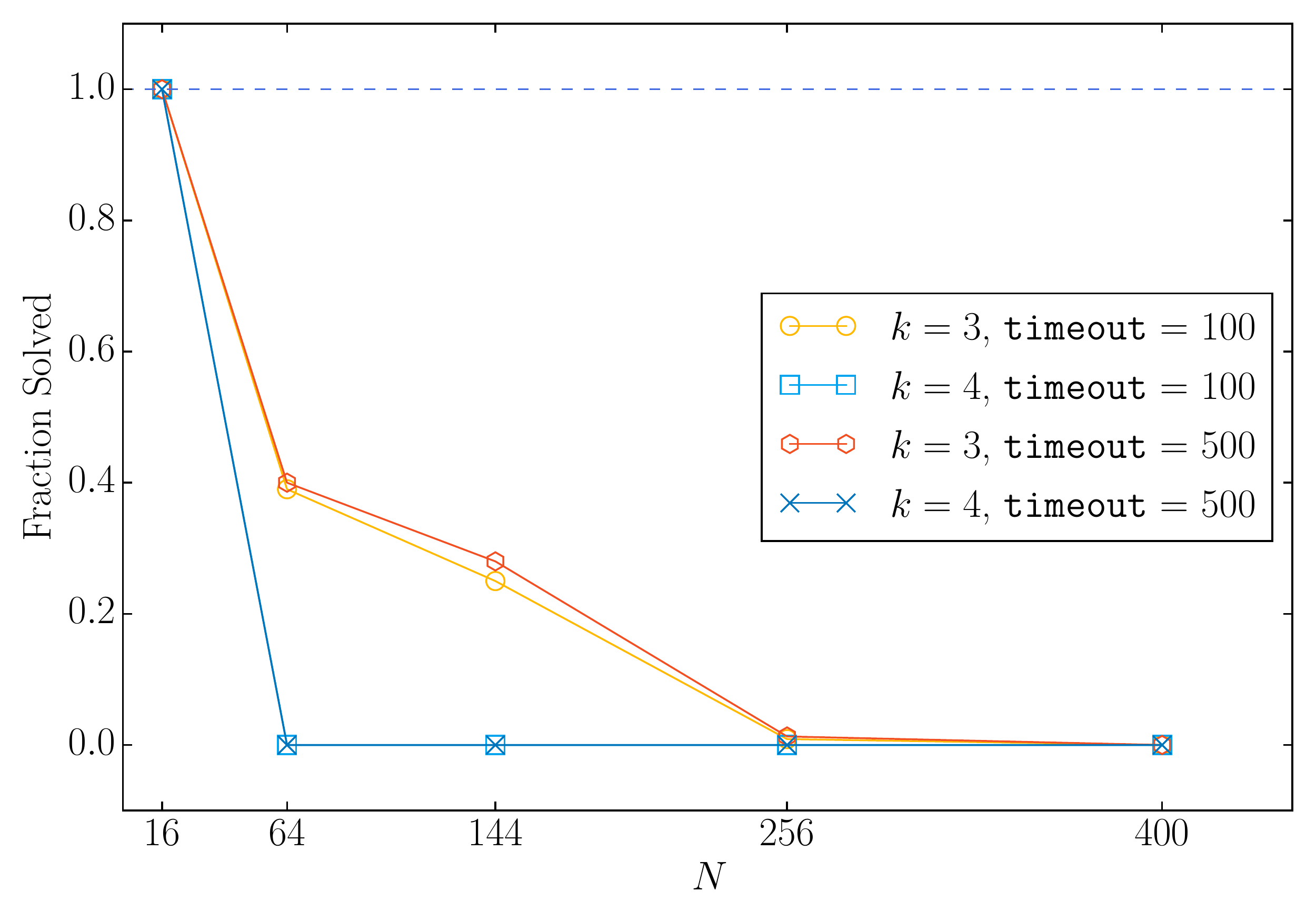}
    \includegraphics[width=\linewidth, keepaspectratio]{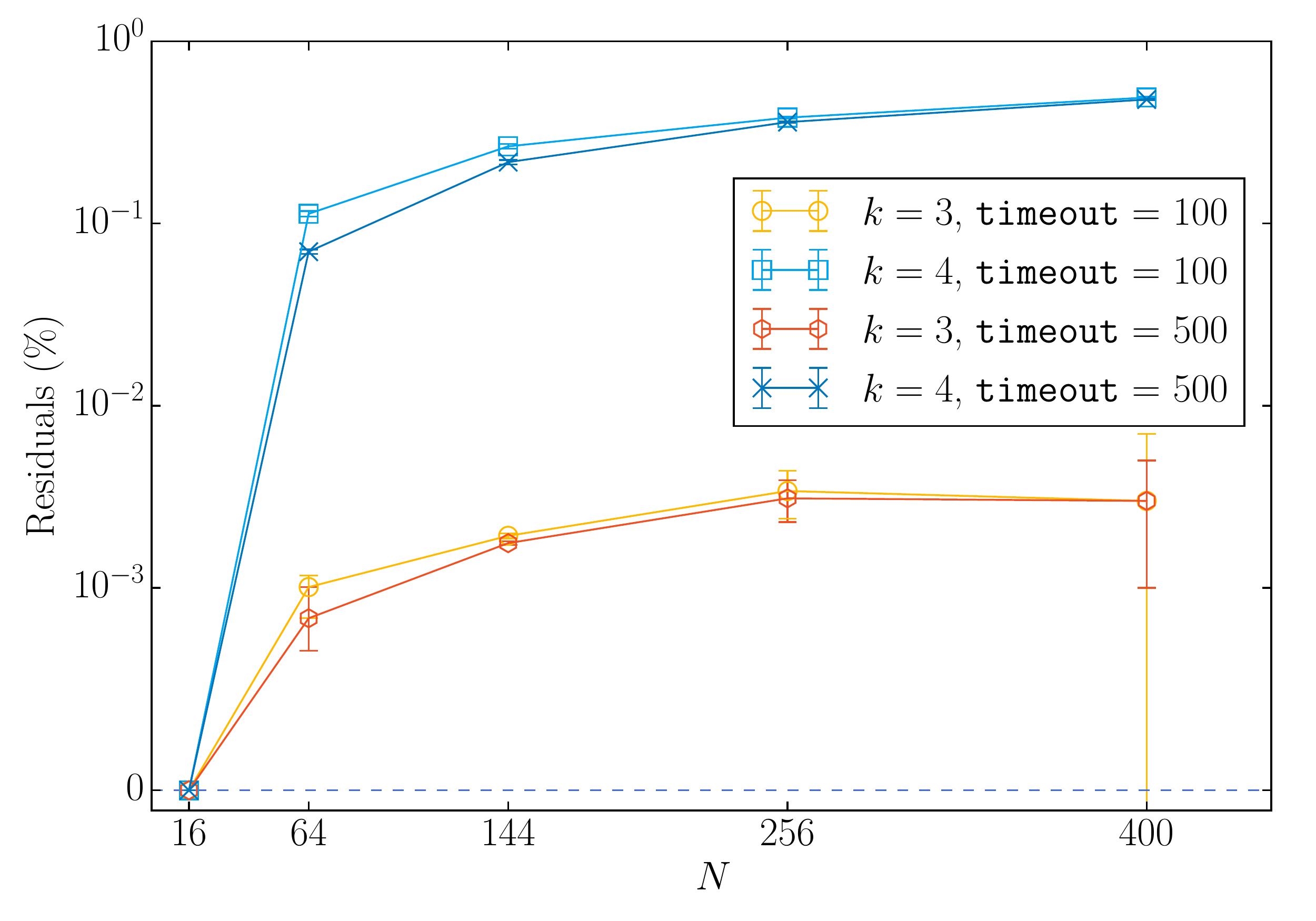}
    \caption{Fraction solved (top) and residuals (bottom) of $2$-local problems obtained by reducing $k$-local instances with $k=3$ and $k=4$. The dashed line represents the reference for the ideal cases. The benchmark experiment has been performed with two different values of the parameter \texttt{timeout}. The data show that solving the $2$-local versions of the problems is extremely difficult. In fact, we were unable to do a scaling analysis as the majority of the problems could not be solved.}
    \label{fig:qubo_results}
\end{figure}

The fraction of solved runs is $100\%$ for all sizes of both $3$- and $4$-local problems. However, it is not possible to calculate the TTS for the $2$-local reductions of either the $3$- or $4$-local problems. Figure~\ref{fig:qubo_results} shows the fraction of solved problems (top panel) and the residual energies (bottom panel). The $2$-local problems derived from the $4$-local instances seem to have a larger overhead than the ones generated from the $3$-local problems, but it is not clear if they have a different computational cost scaling with system size. We surmise that the higher the locality, the larger such overhead in solving the $2$-local reductions will be. The benchmark experiment has been performed with two different values of the parameter \texttt{timeout}. However, increasing the timeout does not improve the quality of the results.

\section{Discussion}
\label{sec:discussion}

\begin{figure*}
    \centering
    \includegraphics[width=0.48\linewidth, keepaspectratio]{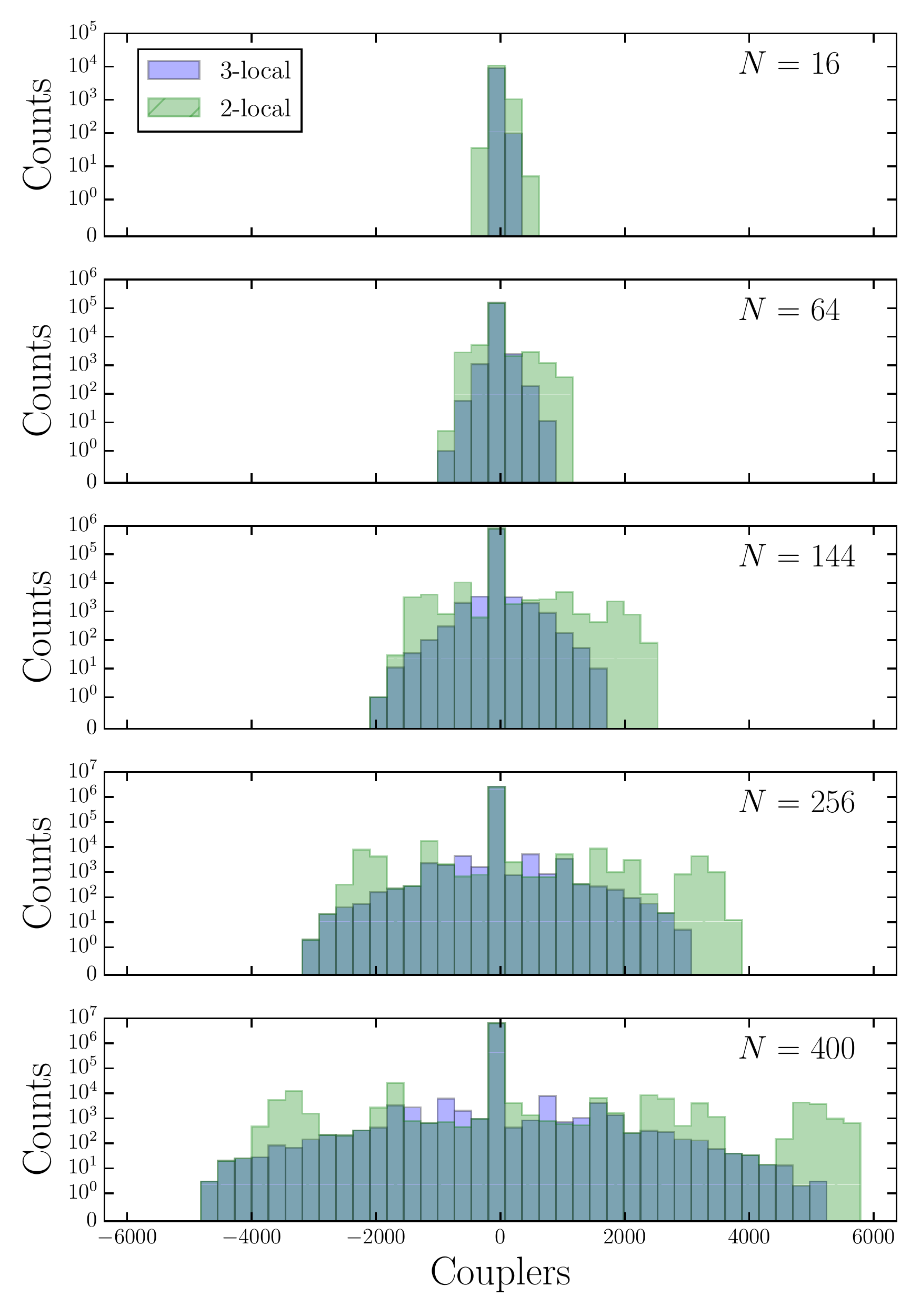}
    \includegraphics[width=0.48\linewidth, keepaspectratio]{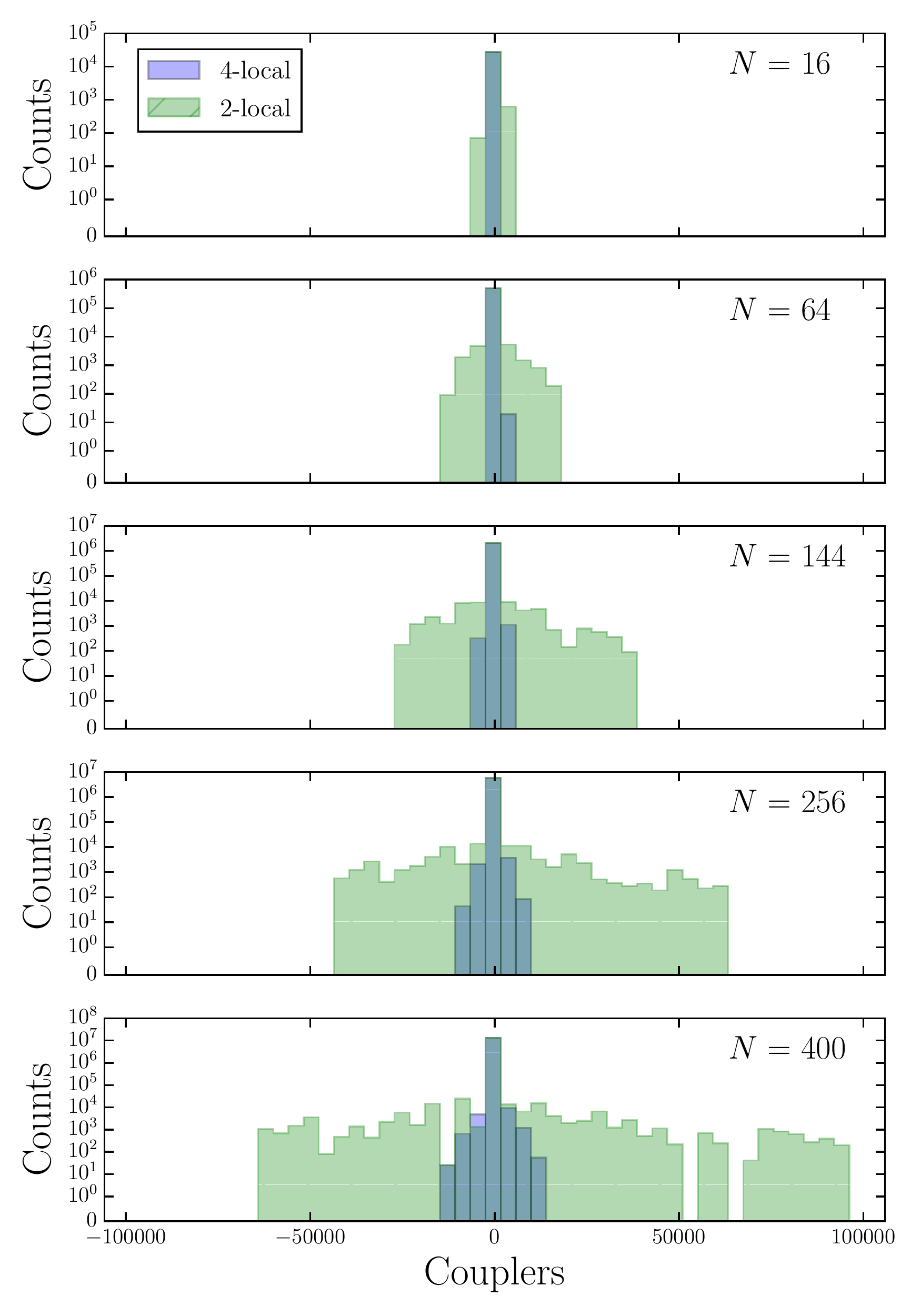}
    \caption{Coupler distributions of $k$-local problems with $k=3$ (left panel) and $k=4$ (right panel) for different system sizes $N$, and their corresponding $2$-local reductions. The distributions of the $3$- and $4$-local problems are quite similar
    (note that the $x$-axis in the plots on the left and the right sides of the figure use a different scale). When comparing the original problems with their reduced form, we observe that, in the $3$-local case, the distributions are comparable, however, weight is redistributed to the tails. In the $4$-local case, there is a sizable increase in the width of the distributions after  locality reduction.}
    \label{fig:couplers_distribution}
\end{figure*}

The computational hardness of the $3$- and $4$-local instances is set in the planting tool \texttt{Chook} by a careful choice of the couplers from different disorder distributions with varying levels of frustration. The reduction to $2$-local interactions in the Hamiltonian requires the introduction of auxiliary variables and penalty terms. Tables~\ref{tb:3l} and \ref{tb:4l} show the increase in the number of variables when reducing the problems to their $2$-local versions. We observe an increase of a factor of approximately $3$ for the \mbox{$3$-local} problems, which increases to a factor $\sim 6$ when reducing the $4$-local problems. Higher-order Hamiltonians will naturally require an even larger overhead. 

\begin{figure}[t!]
    \centering
    \includegraphics[width=\linewidth, keepaspectratio]{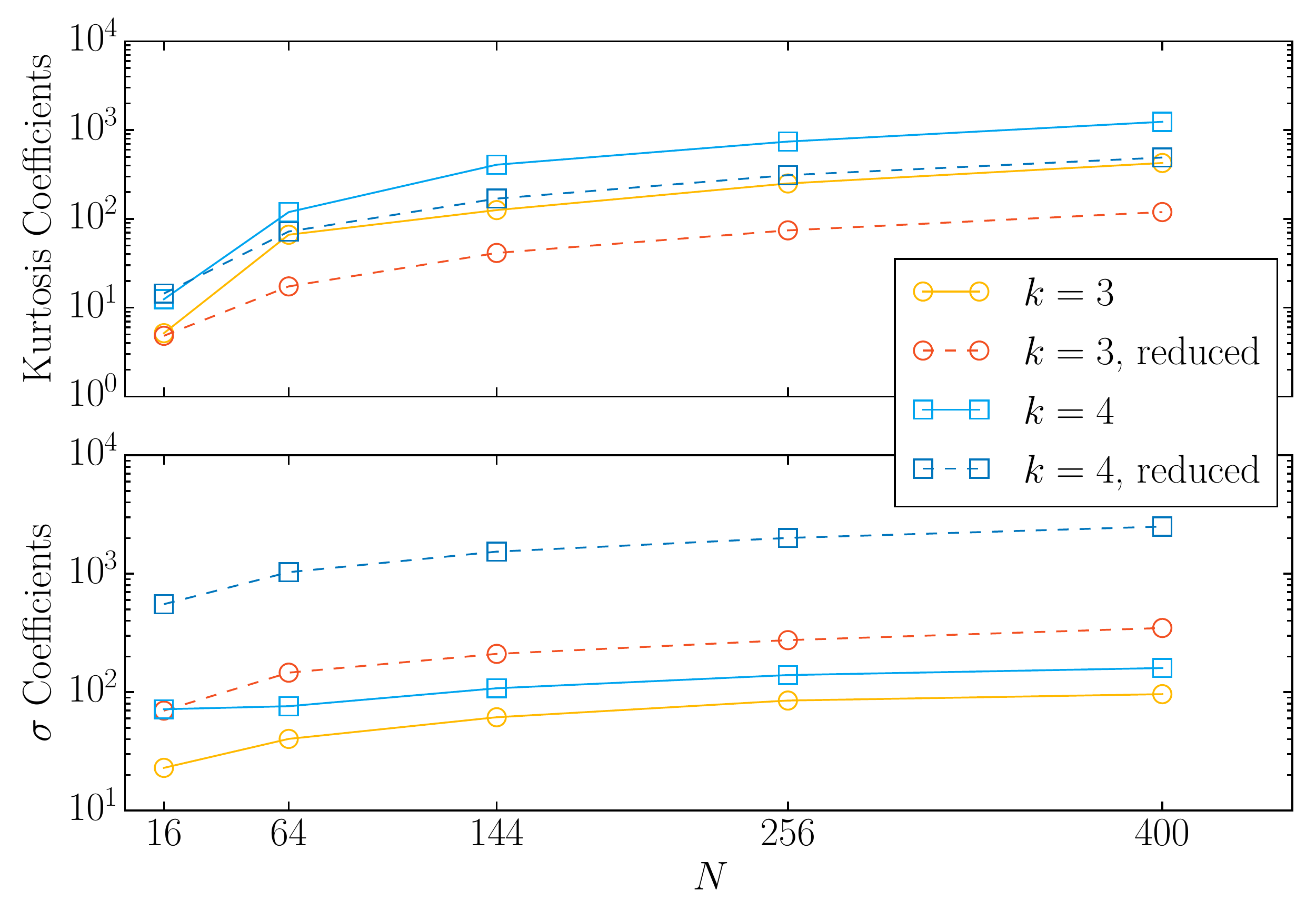}
    \caption{Kurtosis and standard deviation calculated from the coupler distributions of $k$-local problems with $k=3$ and $k=4$, and their correspondent $2$-local reductions. While the kurtosis decreases, the standard deviation of the distributions increases noticeably, thus making the problems harder to solve. Both panels have the same horizontal axis.}
    \label{fig:moments}
\end{figure}

Figure~\ref{fig:couplers_distribution} compares the coupler distributions for the $3$- and \mbox{$4$-local} problems of different system sizes with their corresponding $2$-local reductions. The histograms show that, while the distributions of the $3$- and $4$-local problem are quite similar (note that the $x$-axis in the plots on the left and the right sides of the figure use a different scale), the distributions of their \mbox{$2$-local} reductions are significantly wider, in particular when the reduction occurs from a higher degree of the polynomial, and no longer symmetric. A more quantitative analysis of this effect is shown in Figure~\ref{fig:moments}. We have calculated the standard deviation and the kurtosis of the the coupler distributions. While the former increases by a factor of approximately $10$, the latter reduces by approximately a factor of $5$ when reducing the problems from $k$-local to $2$-local. Having a large dynamic range in the coupler distributions of the reduced problems typically makes these harder to solve with physics-based solvers. 

\begin{figure}[t!]
    \centering
    \includegraphics[width=\linewidth, keepaspectratio]{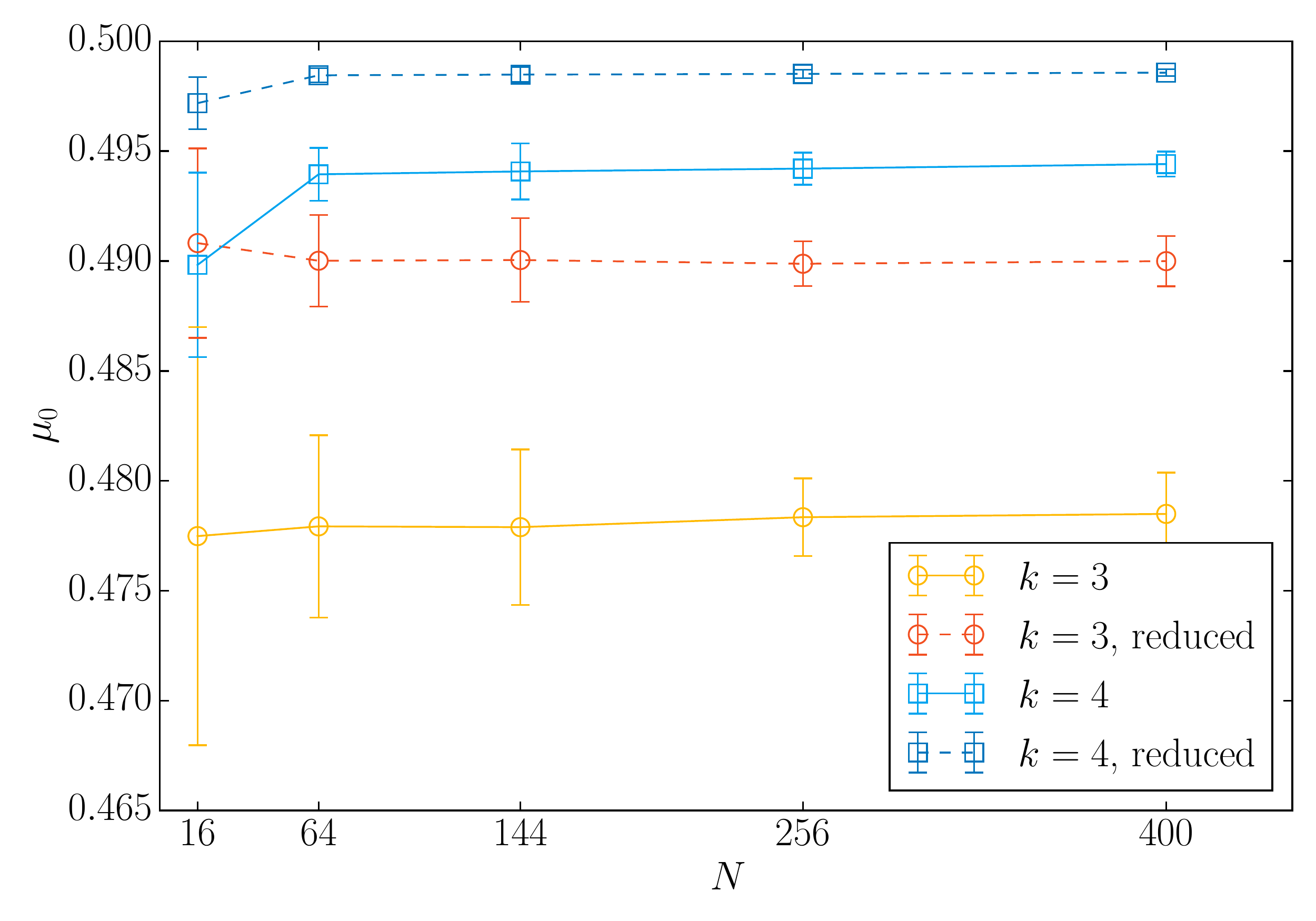}
    \caption{Average misfit parameter $\mu_0$, calculated from the $k$-local instances with $k=3$ and $k=4$, and their correspondent $2$-local reductions. The locality reduction increases the degree of frustration up to $3\%$.}
    \label{fig:misfit}
\end{figure}

The introduction of penalty terms shifts the mean of the coupler distribution by increasing the weight of large positive values: this is an indicator that the frustration will be affected. To confirm this intuition, we have measured the level of frustration for the $3$- and $4$-local instances and their $2$-local reductions. A misfit parameter is used to characterize the degree of frustration of ordered and disordered systems. It measures the increase of the ground-state energy due to frustration, in comparison with that of a relevant reference state. We have measured the level of frustration by calculating the misfit parameter as suggested in \cite{kobe:95}:
\begin{equation}\label{eq:misfit_parameter}
\mu_0 = \frac{E_0-E_{\rm min}^{\rm id}}{E_{\rm max}^{\rm id}-E_{\rm min}^{\rm id}},
\end{equation}
where $E_0$ is the ground energy of our instances, and $E_{\rm min}^{\rm id}$ and $E_{\rm max}^{\rm id}$ describe the minimal and maximal ideally possible energy values, respectively, where ``ideal'' refers to the assumption that all local energies yields a minimal (and maximal) contribution to the total energy. These energies are calculated assuming that all bonds are satisfied (and non-satisfied). Figure~\ref{fig:misfit} presents the average values of the misfit parameter as a function of the problem size calculated over all instances generated for each locality. We observe that the reduction process generates an increase in the degree of frustration level of about $3\%$ for $3$-local instances, and an increase of about $1\%$ for $4$-local instances. The increase in frustration does not show any dependence on the problem size. 

Our conclusion is that the locality reduction makes the problem computationally harder, possibly as a combination of the increase of the number of variables, the greater variance in the coupler distribution, and the change in the frustration level.

\begin{figure*}
    \centering
    \includegraphics[width=\linewidth, keepaspectratio]{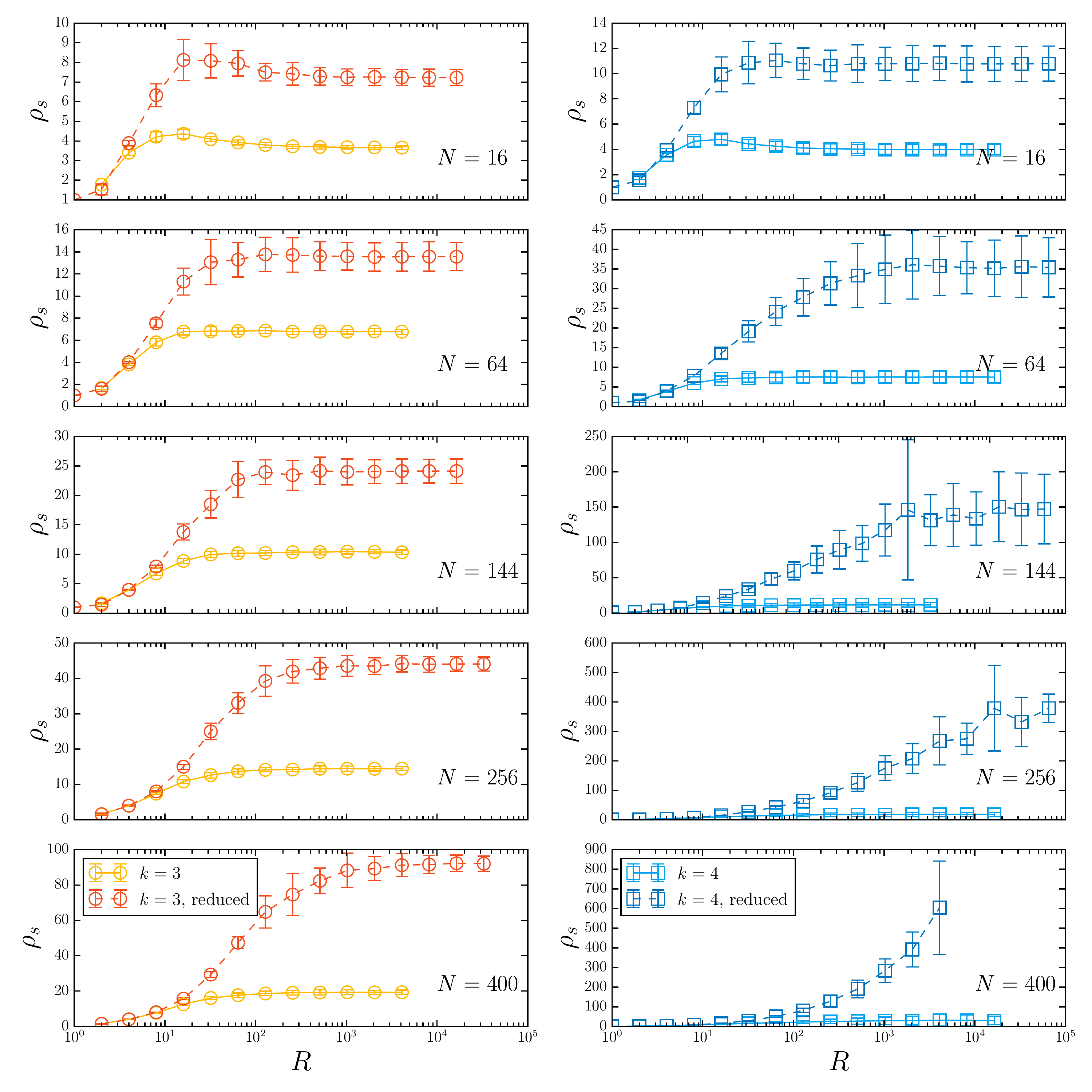}
    \caption{Entropic family size, $\rho_s$, calculated using population annealing Monte Carlo for $k$-local problems with $k=3$ (left panel) and $k=4$ (right panel), for different system sizes $N$, and their corresponding $2$-local reductions. The family size is calculated by averaging over all instances generated for each system size for different values of $R$. The family size of the reduced version of $k=4$ problems with $N=144$ converges only partially, while for larger problem sizes the value does not converge during the allocated timeout.}
    \label{fig:rho_s_R}
\end{figure*}

\begin{figure}
    \centering
    \includegraphics[width=\linewidth, keepaspectratio]{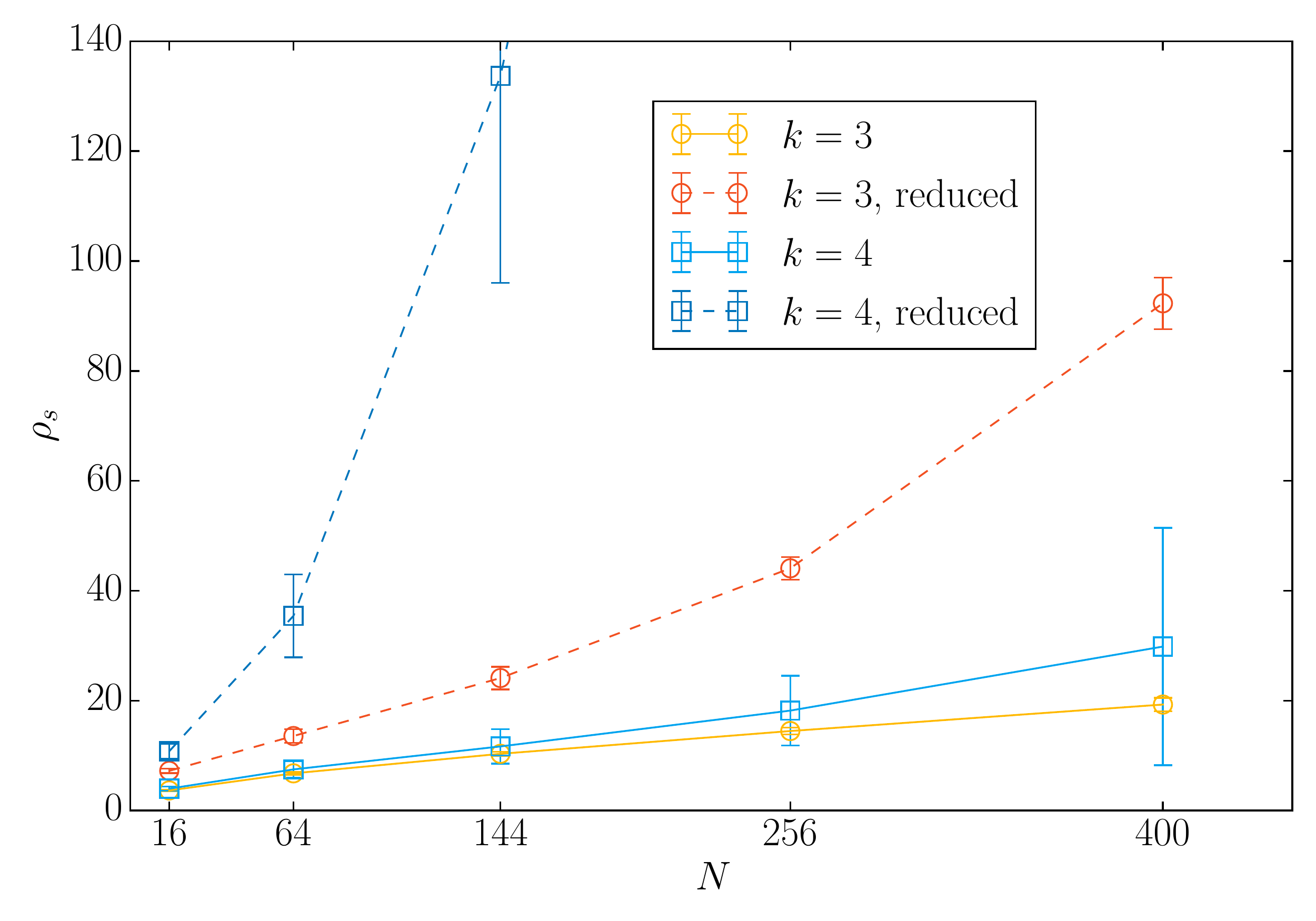}
    \caption{Entropic family size, $\rho_s$, calculated using population annealing Monte Carlo for $k$-local problems with $k=3$ and $k=4$, and their corresponding $2$-local reductions. The family size confirms that a reduction in locality makes the problems computationally harder to solve.}
    \label{fig:rho_s}
\end{figure}

To corroborate the aforementioned observation we use population annealing Monte Carlo (PAMC) \cite{hukushima:03,machta:10,machta:11,wang:15,amey:18,barzegar:18} to measure the {\it entropic family size} $\rho_{\rm s}$. Similar to simulated annealing~(SA)~\cite{kirkpatrick:83}, population annealing is a sequential Markov chain Monte Carlo (MCMC) algorithm in which a population of ``replicas'' is slowly annealed toward a target low temperature. At each temperature, the population is reconfigured via a resampling process during which some replicas are multiplied or eliminated to achieve an equilibrium Gibbs distribution of energies. In a well-thermalized PAMC simulation, a sufficient number of the original replica families must survive. This can be quantified by the {\it family entropy}, $S_{\rm f}$,
\begin{equation}\label{eq:family_entropy}
S_{\rm f} = -\sum_i^R\mathfrak{n}_i\log\mathfrak{n}_i\,,
\end{equation}
where $\mathfrak{n}_i$ is the fraction of the replicas in the $i$-th family and $R$ is the total population size. Large fluctuations in the resampling are a signature of the difficulty in attaining thermal equilibrium, and lead to the descendants of only a few original copies dominating the population~\cite{machta:10}. Hence, a measure of the effective number of surviving replica families at the lowest temperature allows one to distinguish between hard and easy problems. The {\it entropic family size} in thermal equilibrium is defined as
\begin{equation}\label{eq:entropic_family_size}
\rho_{\rm s} = \lim_{R\to\infty}R/e^{S_{\rm f}}.
\end{equation}
For a given set of simulation parameters, the larger the value of $\rho_{\rm s}$, the smaller the number of surviving families, and the more rugged the problem's energy landscape will be. Thus, $\rho_{\rm s}$ provides a measure of hardness for algorithms that are based on local search in the classical energy landscape. As shown in Ref.~\cite{perera:20}, $\rho_{\rm s}$ is highly correlated with other well-established hardness metrics, such as the integrated autocorrelation time in parallel tempering Monte Carlo \cite{wang:15}. Note that $\rho_{\rm s}$, by definition, is an intensive quantity, and therefore independent of the population size $R$ in the thermodynamic limit. In practice, $\rho_{\rm s}$ converges to its true value at a large but finite population.

For all PAMC simulations, we use a linear schedule in inverse temperature. The linear schedule runs between $1/T_{\rm start} = 0$ and $1/T_{\rm end} = 20$ in 100 steps for smaller problems and up to 300 steps for larger ones. In each problem, the coupler values are normalized by the maximum energy scale, such that the same $T_{\rm end}$ can be used for all the benchmarking problems. We perform 10 Metropolis sweeps per replica at each temperature. For all measurements of $\rho_{\rm s}$, we ensure that convergence is achieved unless the simulation times out. We use the following procedure to determine the convergence of the PAMC simulations on each problem instance: starting from a relatively small population size (such as $R=8$), we run PAMC 100 times and record the value of $\rho_{\rm s}$ at the final temperature $T_{\rm end}$ for each restart. The mean value and the corresponding error for each problem are then calculated using the set of computed $\rho_{\rm s}$ values. Then, we double $R$ and repeat the above process. A satisfactory convergence is reached when consecutive values of $\rho_{\rm s}$ agree within the errors. 

Figure~\ref{fig:rho_s_R} shows the mean and standard deviation of $\rho_{\rm s}$ over all instances generated for each problem size, for different values of $R$. The value of $R$ at which $\rho_{\rm s}$ converges depends on the size and hardness of the problem instances and can be visualized as a plateau in the curves. We can observe how $\rho_{\rm s}$ converges to values of $\sim 10$ at a population of $R \simeq 50$ for $3$- and \mbox{$4$-local} problems, while it converges to much higher values for the \mbox{$2$-local} problems. On average, $\rho_{\rm s}$ converges at $R \simeq 10^3-10^4$ for $3$-local reduced instances and $R>10^4$ for \mbox{$4$-local} reduced ones. In particular, measuring $\rho_{\rm s}$ for the reduced version of $4$-local problems is possible for sizes $N=16$ and $N=64$ only. The problems are so hard that the simulation converges only partially for $N=144$ (meaning that not all problem instances converge), and does not converge at all during the allocated time for larger problem sizes. Figure~\ref{fig:rho_s} shows the converging values of $\rho_{\rm s}$ we obtain from the simulation for each system size. The $2$-local reduction critically increases the hardness of the problems, especially for large system sizes. 

Our results demonstrate the advantage of solving the optimization problems in their original $k$-local formulation, and we expect this result to be independent of the choice of solver. Reference~\cite{perdomo_ortiz:19} shows that a simulated quantum annealing~(SQA) algorithm has no advantage in solving a $k$-local formulation of a problem, instead of its $2$-local reduction, but the study includes only instances with $N<20$.

\section{Conclusions}
\label{sec:concl}

We have generated problems with planted solutions having $k$-local interactions and reduced them to their corresponding \mbox{$2$-local} versions, more amenable to current physics-inspired optimization tools than the original ones. The reduction has been performed using a customized version of a classic and extensively adopted quadratization algorithm. The computational time required by the reduction algorithm is known to scale polynomially with the size of the input and thus does not affect the overall exponential scaling found in current physics-inspired optimization methods. Using Microsoft Azure Quantum's implementation of the \texttt{ParallelTempering} parameter-free algorithm, designed to handle problems of any locality, we have attempted to find optima for the native $3$- and $4$-local problems, as well as their $2$-local reductions. All $k$-local problems with $k = 3$ and $k =4$ have been solved to optimality during the allocated \mbox{100-second} timeout. The TTS for \mbox{$4$-local} problems is approximately $5$ times larger than for the $3$-local ones. In contrast, even after increasing the timeout to 500 seconds, the $2$-local reductions could not be solved. It is common practice to apply locality reduction in order to accommodate higher-order polynomial unconstrained optimization problems to run on optimizers that natively handle only quadratic problems, such as the D-Wave quantum annealer, the Fujitsu Digital Annealer, or the Toshiba Simulated Bifurcation Machine. Nevertheless, our results show that doing so should ideally be avoided. As such, investing into creating hardware and/or software to tackle higher-order problems should be prioritized. 

\hspace{2cm}

\section*{Acknowledgements} 
We thank the two anonymous referees for their insightful comments and  valuable suggestions that allowed us to considerably improve our analysis. We thank Marko Bucyk for his careful editing and reviewing of the manuscript. H.~G.~K.~would like to thank David Poulin for inspiring discussions and dedicate this manuscript to him. 


\begin{thebibliography}{10}
\expandafter\ifx\csname url\endcsname\relax
  \def\url#1{\texttt{#1}}\fi
\expandafter\ifx\csname urlprefix\endcsname\relax\def\urlprefix{URL }\fi
\expandafter\ifx\csname href\endcsname\relax
  \def\href#1#2{#2} \def\path#1{#1}\fi

\bibitem{johnson:11}
M.~Johnson, M.~Amin, S.~Gildert, et~al., Quantum annealing with manufactured
  spins, Nature 473 (2011) 194–198.

\bibitem{dickson:13}
N.~Dickson, M.~Johnson, M.~Amin, et~al., Thermally assisted quantum annealing
  of a 16-qubit problem, Nat. Commun. 4~(1903) (2013) 1903.

\bibitem{bunyk:14}
P.~I. Bunyk, E.~M. Hoskinson, M.~W. Johnson, E.~Tolkacheva, F.~Altomare, A.~J.
  Berkley, R.~Harris, J.~P. Hilton, T.~Lanting, A.~J. Przybysz, J.~Whittaker,
  Architectural considerations in the design of a superconducting quantum
  annealing processor, IEEE Trans. Appl. Supercond. 24~(4) (2014) 1–10.

\bibitem{yamaoka:17}
M.~Yamaoka, C.~Yoshimura, M.~Hayashi, T.~Okuyama, H.~Aoki, H.~Mizuno, {A
  20k-Spin Ising Chip to Solve Combinatorial Optimization Problems With CMOS
  Annealing}, IEEE Journal of Solid-State Circuits 51~(1) (2016) 303--309.

\bibitem{yamamotok:17}
K.~Yamamoto, W.~Huang, S.~Takamaeda-Yamazaki, M.~Ikebe, T.~Asai, M.~Motomura,
  {A Time-Division Multiplexing Ising Machine on FPGAs}, Proceedings of the 8th
  International Symposium on Highly Efficient Accelerators and Reconfigurable
  Technologies~(3) (2017) 1--6.

\bibitem{tsukamoto:17}
S.~Tsukamoto, M.~Takatsu, S.~Matsubara, H.~Tamura, {An Accelerator Architecture
  for Combinatorial Optimization Problems}, FUJITSU Sci. Tech. J. 53 (2017)
  8--13.

\bibitem{aramon:19}
M.~Aramon, G.~Rosenberg, E.~Valiante, T.~Miyazawa, H.~Tamura, H.~G. Katzgraber,
  {Physics-Inspired Optimization for Quadratic Unconstrained Problems Using a
  Digital Annealer}, Frontiers in Physics 7 (2019) 48.

\bibitem{okuyama:19}
T.~Okuyama, T.~Sonobe, K.~Kawarabayashi, M.~Yamaoka, {Binary optimization by
  momentum annealing}, Phys. Rev. E 100 (2019) 012111.

\bibitem{patel:20}
S.~Patel, L.~Chen, P.~Canoza, S.~Salahuddin, {Ising Model Optimization Problems
  on a FPGA Accelerated Restricted Boltzmann Machine} (2020).
\newblock \href {http://arxiv.org/abs/2008.04436} {\path{arXiv:2008.04436}}.

\bibitem{yamamotok:21}
K.~Yamamoto, K.~Kawamura, K.~Ando, N.~Mertig, T.~Takemoto, M.~Yamaoka,
  H.~Teramoto, A.~Sakai, S.~Takamaeda-Yamazaki, M.~Motomura, {STATICA: A
  512-Spin 0.25M-Weight Annealing Processor With an All-Spin-Updates-at-Once
  Architecture for Combinatorial Optimization With Complete Spin–Spin
  Interactions}, IEEE Journal of Solid-State Circuits 56~(1) (2021) 165--178.

\bibitem{leleu:21}
T.~Leleu, F.~Khoyratee, T.~Levi, R.~Hamerly, T.~Kohno, K.~Aihara, {Scaling
  advantage of nonrelaxational dynamics for high-performance combinatorial
  optimization} (2021).
\newblock \href {http://arxiv.org/abs/2009.04084} {\path{arXiv:2009.04084}}.

\bibitem{wang:13}
Z.~Wang, A.~Marandi, K.~Wen, R.~L. Byer, Y.~Yamamoto, {Coherent Ising machine
  based on degenerate optical parametric oscillators}, Phys. Rev. A 88 (2013)
  063853.

\bibitem{marandi:14}
A.~Marandi, Z.~Wang, K.~Takata, R.~L. Byer, Y.~Yamamoto, {Network of
  time-multiplexed optical parametric oscillators as a coherent Ising machine},
  Nat. Photonics 8~(12) (2014) 937–942.

\bibitem{inagaki:16}
T.~Inagaki, Y.~Haribara, K.~Igarashi, T.~Sonobe, S.~Tamate, T.~Honjo,
  A.~Marandi, P.~L. McMahon, T.~Umeki, K.~Enbutsu, O.~Tadanaga, H.~Takenouchi,
  K.~Aihara, K.-i. Kawarabayashi, K.~Inoue, S.~Utsunomiya, H.~Takesue, {A
  coherent Ising machine for 2000-node optimization problems}, Science
  354~(6312) (2016) 603--606.

\bibitem{mcmahon:16}
P.~L. McMahon, A.~Marandi, Y.~Haribara, R.~Hamerly, C.~Langrock, S.~Tamate,
  T.~Inagaki, H.~Takesue, S.~Utsunomiya, K.~Aihara, R.~L. Byer, M.~M. Fejer,
  H.~Mabuchi, Y.~Yamamoto, {A fully programmable 100-spin coherent Ising
  machine with all-to-all connections}, Science 354~(6312) (2016) 614--617.

\bibitem{yamamotoy:17}
Y.~Yamamoto, K.~Aihara, T.~Leleu, K. Kawarabayashi, S.~Kako, M.~Fejer,
  K.~Inoue, H.~Takesue, {Coherent Ising machines---optical neural networks
  operating at the quantum limit}, npj Quantum Inf. 3~(1) (2017) 49.

\bibitem{hamerly:19}
R.~Hamerly, T.~Inagaki, P.~L. McMahon, D.~Venturelli, A.~Marandi, T.~Onodera,
  E.~Ng, C.~Langrock, K.~Inaba, T.~Honjo, K.~Enbutsu, T.~Umeki, R.~Kasahara,
  S.~Utsunomiya, S.~Kako, K.-i. Kawarabayashi, R.~L. Byer, M.~M. Fejer,
  H.~Mabuchi, D.~Englund, E.~Rieffel, H.~Takesue, Y.~Yamamoto, {Experimental
  investigation of performance differences between coherent Ising machines and
  a quantum annealer}, Science Advances 5~(5) (2019).

\bibitem{pierangeli:19}
D.~Pierangeli, G.~Marcucci, C.~Conti, {Large-Scale Photonic Ising Machine by
  Spatial Light Modulation}, Phys. Rev. Lett. 122~(21) (2019) 213902.

\bibitem{pierangeli:20a}
D.~Pierangeli, G.~Marcucci, D.~Brunner, C.~Conti, {Noise-enhanced
  spatial-photonic Ising machine}, Nanophotonics 9~(13) (2020) 4109–4116.

\bibitem{pierangeli:20b}
D.~Pierangeli, G.~Marcucci, C.~Conti, {Adiabatic evolution on a
  spatial-photonic Ising machine}, Optica 7~(11) (2020) 1535--1543.

\bibitem{pierangeli:21}
D.~Pierangeli, M.~Rafayelyan, C.~Conti, S.~Gigan, {Scalable Spin-Glass Optical
  Simulator}, Phys. Rev. Applied 15 (2021) 034087.

\bibitem{goto:19b}
H.~Goto, K.~Tatsumura, A.~R. Dixon, {Combinatorial optimization by simulating
  adiabatic bifurcations in nonlinear Hamiltonian systems}, Science Advances
  5~(4) (2019) eaav2372.

\bibitem{goto:21}
H.~Goto, K.~Endo, M.~Suzuki, Y.~Sakai, T.~Kanao, Y.~Hamakawa, R.~Hidaka,
  M.~Yamasaki, K.~Tatsumura, {High-performance combinatorial optimization based
  on classical mechanics}, Science Advances 7~(6) (2021) eabe7953.

\bibitem{goto:16}
H.~Goto, {Bifurcation-based adiabatic quantum computation with a nonlinear
  oscillator network}, Sci. Rep. 6~(1) (2016) 21686.

\bibitem{nigg:17}
S.~E. Nigg, N.~L{\"o}rch, R.~P. Tiwari, {Robust quantum optimizer with full
  connectivity}, Science Advances 3~(4) (2017) e1602273.

\bibitem{puri:17}
S.~Puri, C.~K. Andersen, A.~L. Grimsmo, A.~Blais, {Quantum annealing with
  all-to-all connected nonlinear oscillators}, Nat. Commun. 8~(1) (2017) 15785.

\bibitem{goto:18}
H.~Goto, Z.~Lin, Y.~Nakamura, {Boltzmann sampling from the Ising model using
  quantum heating of coupled nonlinear oscillators}, Sci. Rep. 8~(1) (2018)
  7154.

\bibitem{goto:19a}
H.~Goto, {Quantum Computation Based on Quantum Adiabatic Bifurcations of
  Kerr-Nonlinear Parametric Oscillators}, J. Phys. Soc. Jpn. 88~(6) (2019)
  061015.

\bibitem{perdomo_ortiz:19}
A.~Perdomo-Ortiz, A.~Feldman, A.~Ozaeta, S.~V. Isakov, Z.~Zhu, B.~O’Gorman,
  H.~G. Katzgraber, A.~Diedrich, H.~Neven, J.~de~Kleer, et~al., {Readiness of
  Quantum Optimization Machines for Industrial Applications}, Phys. Rev.
  Applied 12~(1) (2019) 014004.

\bibitem{konz:21}
M.~S. K{\"o}nz, W.~Lechner, H.~G. Katzgraber, M.~Troyer, {Scaling overhead of
  embedding optimization problems in quantum annealing} (2021).
\newblock \href {http://arxiv.org/abs/2103.15991} {\path{arXiv:2103.15991}}.

\bibitem{zhu:16}
Z.~Zhu, A.~J. Ochoa, S.~Schnabel, F.~Hamze, H.~G. Katzgraber, Best-case
  performance of quantum annealers on native spin-glass benchmarks: How chaos
  can affect success probabilities, Phys. Rev. A 93 (2016) 012317.

\bibitem{albash:19}
T.~Albash, V.~Martin-Mayor, I.~Hen, Analog errors in {Ising} machines, Quantum
  Sci. Technol. 4~(2) (2019) 02LT03.

\bibitem{de_las_cuevas:09}
G.~D. las Cuevas, W.~D{\"u}r, M.~V. den Nest, H.~J. Briegel, {Completeness of
  classical spin models and universal quantum computation}, J. Stat. Mech.
  2009~(07) (2009) P07001.

\bibitem{de_las_cuevas:10}
G.~D. las Cuevas, W.~Dür, H.~J. Briegel, M.~A. Martin-Delgado, {Mapping all
  classical spin models to a lattice gauge theory}, New J. Phys. 12~(4) (2010)
  043014.

\bibitem{andrist:11}
R.~S. Andrist, H.~G. Katzgraber, H.~Bombin, M.~A. Martin-Delgado, {Tricolored
  lattice gauge theory with randomness: fault tolerance in topological color
  codes}, New J. Phys. 13~(8) (2011) 083006.

\bibitem{feldman:10}
A.~Feldman, G.~Provan, A.~Van~Gemund, {Approximate Model-Based Diagnosis Using
  Greedy Stochastic Search}, Journal of Artificial Intelligence Research 38
  (2010) 371–413.

\bibitem{hernandez:16}
M.~Hernandez, A.~Zaribafiyan, M.~Aramon, M.~Naghibi, {A Novel Graph-Based
  Approach for Determining Molecular Similarity} (2016).
\newblock \href {http://arxiv.org/abs/1601.06693} {\path{arXiv:1601.06693}}.

\bibitem{marchand:19}
D.~J.~J. Marchand, M.~Noori, A.~Roberts, G.~Rosenberg, B.~Woods, U.~Yildiz,
  M.~Coons, D.~Devore, P.~Margl, {A Variable Neighbourhood Descent Heuristic
  for Conformational Search Using a Quantum Annealer}, Sci. Rep. 9 (2019)
  13708.

\bibitem{qmedia}
{Microsoft Quantum}, {Jij and Toyota Tsusho: reducing carbon emissions with
  Azure Quantum},
  \url{https://cloudblogs.microsoft.com/quantum/2020/08/04/jij-toyota-azure-quantum-reducing-carbon-emissions/}, accessed: 2021-07-19.

\bibitem{boros:14}
E.~Boros, A.~Gruber, {On Quadratization of Pseudo-Boolean Functions} (2014).
\newblock \href {http://arxiv.org/abs/1404.6538} {\path{arXiv:1404.6538}}.

\bibitem{boros:02}
E.~Boros, P.~L. Hammer, {Pseudo-Boolean optimization}, Discrete Applied
  Mathematics 123~(1) (2002) 155.

\bibitem{dattani:19}
N.~Dattani, {Quadratization in discrete optimization and quantum mechanics}
  (2019).
\newblock \href {http://arxiv.org/abs/1901.04405} {\path{arXiv:1901.04405}}.

\bibitem{perera:20chook}
D.~Perera, I.~Akpabio, F.~Hamze, S.~Mandr\`{a}, N.~Rose, M.~Aramon, H.~G.
  Katzgraber, {Chook -- A comprehensive suite for generating binary
  optimization problems with planted solutions} (2020).
\newblock \href {http://arxiv.org/abs/2005.14344} {\path{arXiv:2005.14344}}.

\bibitem{hamze:18}
F.~Hamze, D.~C. Jacob, A.~J. Ochoa, D.~Perera, W.~Wang, H.~G. Katzgraber, From
  near to eternity: Spin-glass planting, tiling puzzles, and
  constraint-satisfaction problems, Phys. Rev. E 97~(4) (Apr 2018).

\bibitem{perera:20}
D.~Perera, F.~Hamze, J.~Raymond, M.~Weigel, H.~G. Katzgraber, Computational
  hardness of spin-glass problems with tile-planted solutions, Phys. Rev. E
  101~(2) (Feb 2020).

\bibitem{rosenberg:75}
I.~Rosenberg, {Reduction of bivalent maximization to the quadratic case},
  Cahiers du Centre d'Etudes de Recherche Operationnelle 17 (1975) 71.

\bibitem{1qloud}
1QBit, {1Qloud Documentation: Convert HOBO to QUBO},
  \url{https://portal.1qbit-prod.com/docs/task/convert-hobo-to-a-qubo},
  accessed: 2020-12-09.

\bibitem{pt}
{Microsoft Quantum}, {Parallel Tempering},
  \url{https://docs.microsoft.com/en-ca/azure/quantum/optimization-parallel-tempering},
  accessed: 2021-05-14.

\bibitem{ronnow:14}
T.~F. R{\o}nnow, Z.~Wang, J.~Job, S.~Boixo, S.~V. Isakov, D.~Wecker, J.~M.
  Martinis, D.~A. Lidar, M.~Troyer, Defining and detecting quantum speedup,
  Science 345~(6195) (2014) 420–424.

\bibitem{kobe:95}
S.~Kobe, T.~Klotz, Frustration: How it can be measured, Phys. Rev. E 52~(5)
  (1995) 5660–5663.

\bibitem{hukushima:03}
K.~Hukushima, Y.~Iba, {Population Annealing and Its Application to a Spin
  Glass}, in: J.~E. Gubernatis (Ed.), {The Monte Carlo method in the physical
  sciences: celebrating the 50th anniversary of the Metropolis algorithm}, Vol.
  690, AIP, Los Alamos, New Mexico (USA), 2003, p. 200.

\bibitem{machta:10}
J.~Machta, {Population annealing with weighted averages: A Monte Carlo method
  for rough free-energy landscapes}, Phys. Rev. E 82 (2010) 026704.

\bibitem{machta:11}
J.~Machta, R.~Ellis, {Monte Carlo Methods for Rough Free Energy Landscapes:
  Population Annealing and Parallel Tempering}, J. Stat. Phys. 144 (2011) 541.

\bibitem{wang:15}
W.~Wang, J.~Machta, H.~G. Katzgraber, {Population annealing: Theory and
  application in spin glasses}, Phys. Rev. E 92 (2015) 063307.

\bibitem{amey:18}
C.~Amey, J.~Machta, {Analysis and optimization of population annealing}, Phys.
  Rev. E 97 (2018) 033301.

\bibitem{barzegar:18}
A.~Barzegar, C.~Pattison, W.~Wang, H.~G. Katzgraber, {Optimization of
  population annealing Monte Carlo for large-scale spin-glass simulations},
  Phys. Rev. E 98 (2018) 053308.

\bibitem{kirkpatrick:83}
S.~Kirkpatrick, C.~D. Gelatt, Jr., M.~Vecchi, {Optimization by Simulated
  Annealing}, Science 220 (1983) 671.

\end{thebibliography}

\end{document}